\documentclass[12pt,a4paper]{article}

\usepackage{amsmath}
\usepackage{multirow}

\usepackage{graphicx}

\title{Optimization of factorization scale in QED Drell-Yan-like processes}

\author{Andrej Arbuzov$^{1,2}$,
  Uliana Voznaya$^{1,2}$,
  Aliaksandr Sadouski$^{3,4}$}

\date{}

\begin{document}

\sloppypar

\maketitle

\begin{center} {\it
$^1$ Joint Institute for Nuclear Research, Joliot-Curie str. 6, Dubna, 141980, Russia \\
$^2$ Dubna state university, Universitetskaya str. 19, Dubna, 141980, Russia \\
$^3$ Gomel State Medical University,
246000, Lange str. 5, Gomel, Belarus \\
$^4$ Francisk Skorina Gomel State University,
246019, Sovetskaya str. 104, Gomel, Belarus
}
\end{center}


\begin{abstract}
  The dependence of corrections due to the initial state radiation in $e^+ e^-$-annihilation processes on the choice of the factorization scale is investigated.
Different prescriptions of the factorization scale choice are analyzed within the leading and next-to-leading logarithmic approximations. Comparisons with the known complete one-loop and two-loop results are used to optimize the scale choice.  
\end{abstract}

\section{Introduction}
Current and future high-precision experiments challenge the theory by demanding very accurate predictions which require in particular calculation of higher-order contributions within perturbation theory. Precise theoretical predictions of radiative corrections to electron-positron annihilation are important for experiments at future electron-positron colliders including FCC-ee \cite{FCC:2018evy} and CEPC~\cite{CEPCStudyGroup:2018ghi}.
Despite the great progress in the development of multi-loop computational methods, computing the full $\mathcal{O}(\alpha^2)$ electroweak and even QED radiative corrections to realistic observables in high-energy physics remains a complex task. The QED structure function\footnote{These functions are better to be called QED parton distribution functions
(QED PDFs).} approach~\cite{Kuraev:1985hb} circumvents this problem by systematically taking into account higher-order contributions, amplified by the powers of the so-called large logarithm
\begin{equation}
L = \ln \frac{\mu_F^2}{\mu_R^2},
\end{equation}
where $\mu_F$ is the factorization scale and $\mu_R$ is the renormalization scale. Here, we consider the perturbative expansion of radiative corrections in the series both in $\alpha$
and the large logarithm. Terms of the order $\mathcal{O}(\alpha^n L^n)$ are usually called the leading logarithmic contributions (LL), and those of the order $\mathcal{O}(\alpha^nL^{n-1})$ are the next-to-leading logarithmic (NLL) ones, and so on. 
Recently NNLL electron and photon PDFs were calculated in the ${\mathcal O}(\alpha^2)$ order~\cite{Stahlhofen:2025hqd,Schnubel:2025ejl}.

The renormalization scale in QED is usually chosen to be equal to the electron mass $m_e$, 
so in what follows we will use $\alpha(\mu_R^2)\equiv\alpha\approx\alpha(0)$. 
This choice is dictated by the need to explicitly define and control the terms singular in the electron mass. 
Regarding the factorization scale, its choice is somewhat arbitrary. 
The known explicit analytic results show that the factorization scale should correspond 
to the characteristic energy scale of the process under consideration, e.g., 
the factorization scale equal to the $t$-channel momentum transfer $\mu_F=\sqrt{-t}$ 
is a reasonable choice for small-angle Bhabha scattering~\cite{Arbuzov:2006mu}. 
One should keep in mind that the ultimate result of calculations of a (renormalization invariant) 
observable quantity in all orders in $\alpha$ and $L$ should not depend on the choice of both 
renormalization and factorization scales. Moreover, for any given order in $\alpha$ the sum of all 
contributions with different powers of $L$ should not depend on the factorization scale value.
Variation of the factorization scale leads to redistribution of contributions between 
terms with different powers of $L$ in horizontal lines of the following formal representation 
of radiative corrections 

\begin{align} \label{triangle}
  \begin{tabular}{l c c c c}
   &  $\alpha^0~(\mathrm{Born})$  & & &  \\ 
   &  $\alpha^1L^1$ \quad \ \ & $\alpha^1L^0$ & &  \\ 
   & $\alpha^2L^2$ \quad \ \ & $\alpha^2L^1$ &\qquad $\alpha^2L^0$ & \\ 
   & $\alpha^3L^3$ \quad \ \ & $\alpha^3L^2$ &\qquad $\alpha^3L^1$ & $\alpha^3L^0$ \\ 
   & $\ldots$ \quad \ \ & & & 
    \end{tabular}  
\end{align}
Optimization of the scale choice should help to 
concentrate most of the corrections in terms with higher powers of the large logarithms, i.e., the LL and NLL ones. At the same time, we try to reduce the contribution of the terms with lower powers of $L$ (and higher powers in $\alpha$) which are more difficult to calculate. In this way, we attempt to improve the convergence of perturbative calculations and reduce the theoretical uncertainty due to unknown higher-order terms. 


Here, we consider the specific case of electron-positron annihilation into a virtual photon or $Z$ boson.
This process can be treated exactly in the same way as the Drell-Yan process in QCD with factorization 
between parton density functions (PDFs) in the initial state and the hard subprocess, 
see ref.~\cite{Arbuzov:2024tac} for details. We now have a unique opportunity to analyze quantitatively the structure of representation~(\ref{triangle}), since the $\mathcal{O}(\alpha^2L^0)$ contribution and many high-order LL and NLL terms have been calculated analytically. Note that there is practically no any other high-energy process for which we have an analytically known differential distribution with such detail.

The article is organized as follows. In the next Section we discuss the application of several approaches 
for choosing the factorization scale. The Section is divided into three parts. In the first part, 
we compare the differential distributions obtained with three different factorization scales. 
In the second part, we analyze the effects of varying the factorization scale in the total cross-section. 
In the third part, we test the effectiveness of the standard scale variation procedure 
($\mu_F \to \mu_F/2,\ 2 \mu_F$), commonly used to estimate the uncertainties due to the scale dependence. 
The final Section contains concluding remarks.

In this article we systematically compare $\mu = \sqrt{s}$, $\mu = \sqrt{sz}$, and $\mu = \sqrt{s/e}$, where $e \approx 2.71828$ is Euler's number, make a quantitative test of LL and NLL
approximations against the known complete $\mathcal{O}(\alpha^2)$ result. We study also the reliability of
the conventional factor-of-two factorization-scale variation and observe that this variation can underestimate the uncertainty if radiative return to lower energies is enhanced.

\section{\label{sec:level1}Factorization scale choices}

There are known many ways to optimize the factorization scale choice, see e.g.~\cite{DiGiustino:2022ggl}. 
In the conventional scale setting (CSS) scheme, $\mu_F$ is chosen to be equal to the characteristic energy
scale of the hard sub-process, for example, to the invariant mass of the final state lepton pair in 
Drell-Yan-like processes at the LHC~\cite{Anastasiou:2003ds} and to the muon mass for the muon decay 
spectrum~\cite{Arbuzov:2002pp}. However, this approach has a number of drawbacks: in some processes there 
can be more than one characteristic scale (e.g., in large-angle Bhabha scattering we have amplitudes 
in $s$, $t$, and $u$ channels), the results can depend also on the choice of the renormalization scale and scheme.

The method of the fastest apparent convergence (FAC) \cite{Grunberg:1982fw} exploits the notion of effective
charges. A similar approach was also proposed by N.V.~Krasnikov~\cite{Krasnikov:1981rp}. The basic idea 
is that an effective charge can be associated with a given physical quantity, and higher-order corrections 
to the quantity can be absorbed by renormalization of that charge, so that the LL-corrected expression 
is equal to the sum of the LL and NLL corrections~\cite{Ingelman:1994nz}. In the case of differential
distributions and if anomalous dimensions contribute to the physical quantity of interest, 
the FAC method suggests choosing the $\mu_F$ value which corresponds to the fastest convergence 
of the series of the results obtained in the LL, NLL, NNLL, and further approximations. 
 
The Brodsky-Lepage-Mackenzie (BLM) \cite{Brodsky:1982gc} procedure is based on the resummation of vacuum
polarization effects in the running coupling constant. The optimal renormalization and factorization scales
for a particular process at a given order can be determined by computing the vacuum polarization parts 
of the diagrams of that order, and by setting the scale with which the terms proportional to the number 
of flavors $N_f$ are absorbed into the coupling constant. The principle of maximal conformality 
(PMC)~\cite{Brodsky:2011zza,Yan:2023hra} is a generalization of the BLM prescription to all orders and 
different types of observables. This approach aims to make results of perturbative calculation being invariant 
with respect to the renormalization scale and scheme choices. The main idea is to exploit the renormalization 
group invariance. As a result, the convergence of the perturbative QCD series can be improved. It should also 
be noted that the application of the PMC approach requires careful consideration of higher-order 
effects~\cite{Chawdhry:2019uuv,Kataev:2014jba,Kataev:2023xqr}.

The principle of minimal sensitivity (PMS) was proposed by P.M.~Stevenson~\cite{Stevenson:2022gcv}. 
According to this principle, the optimal factorization scale is the one for which the result is 
least sensitive to the value of the scale. According to PMS, the scale can be chosen by equating 
the derivative of a physical quantity of interest with respect to the factorization scale to zero 
(or to its apparent minimum value).

Our goal here is to analyze the applicability and effectiveness of various scale selection prescriptions 
for a specific QED process where analytical results are known for several orders in $\alpha$ and $L$.

\subsection{Factorization scale choices in differential distribution}

Let's consider the process of electron-positron annihilation 
\begin{eqnarray}
  && e^-(p_1) + e^+(p_2) \to \gamma^*,Z \to \mu^-(q_1) + \mu^+(q_2) + (n\gamma,e^+e^-),
  \nonumber \\
  && s = (p_1+p_2)^2, \qquad s'=(q_1+q_2)^2= zs,
\end{eqnarray}
where $(n \gamma,e^+e^-)$ means possible emitted photons and/or $e^+e^-$ pairs. 

The contribution to the cross section of this process due to one-loop initial state radiation (ISR) corrections reads~\cite{Berends:1987ab}  
\begin{equation}
\frac{d\sigma^{(1)}_{\bar{e} e}(s')}{ds'} =   \frac{\alpha}{\pi}\frac{\sigma^{(0)}(s')}{s} \biggl\{
\left[\frac{1+z^2}{1-z}\right]_+\left(\ln\frac{s}{m_e^2}-1\right) + \delta(1-z)
\left(2\zeta(2)-\frac{1}{2}\right)\biggr\}
+ \mathcal{O}\left(\frac{m_e^2}{s}\right), 
\end{equation}
where the standard plus prescription is applied and $\zeta(2)=\pi^2/6$.
The corresponding complete two-loop result was presented in~\cite{Berends:1987ab} and corrected in~\cite{Blumlein:2019srk}.
These results can be compared with those obtained by means of
the QED structure function approach~\cite{Kuraev:1985hb} within the next-to-leading 
logarithmic approximation~\cite{Arbuzov:2024tac,Ablinger:2020qvo}
where the cross section is represented in the Drell-Yan form
\begin{eqnarray} \label{master}
&&	\frac{d\sigma^{\mathrm{NNLL}}_{\bar{e} e}(s')}{ds'} = \sum \limits_{i,j= e\!, \bar{e}\!,  \gamma}  \int \limits^{1}_{\bar{z}_1} \int \limits^{1}_{\bar{z}_2} d z_1 
	d z_2 D^{\mathrm{str}}_{i e} \left(z_1,\frac{\mu_R^2}{\mu^2_F}\right)  D^{\mathrm{str}}_{j \bar{e}} \! \left( \! z_2,\frac{\mu_R^2}{\mu^2_F} \! \right) \!
	\nonumber \\
&&	\times
	\left( \sigma^{(0)}_{ij} (s z_1 z_2) + \bar{\sigma}^{(1)}_{ij} (s z_1 z_2) +
    \bar{\sigma}^{(2)}_{ij} (s z_1 z_2) 
    \right)
    \delta(s' \! - \! sz_1z_2)
	+ \mathcal{O}\left(\frac{m_e^2}{s}\right),
\end{eqnarray}
where $D^{\mathrm{str}}_{ie}(x,\mu_R^2/\mu^2_F)$ are the electron structure (parton density) functions~\cite{Kuraev:1985hb,Arbuzov:2022fmv,Bertone:2019hks}. The lower limits $\bar{z}_{1,2}$ of integration over energy fractions depend on experimental conditions. For $D \otimes D\otimes\sigma^{(0)}$ 
$z \equiv z_1z_2$, and $sz$ is the square of the observed invariant mass of the muon pair in the final state.
Expansion of this master formula in $\alpha$ gives in the first order
\begin{eqnarray}
\label{matchingOa1}
 \frac{d\sigma^{(1)}_{\bar{e} e}(s')}{ds'} = 2 \frac{\alpha}{2\pi}\left[ P^{(0)}_{ee}(z) \ln \frac{\mu_F^2}{\mu_R^2} + d_{ee}^{(1)}(z)\right]  \frac{\sigma^{(0)}_{\bar{e} e}(s')}{s} +  \frac{\bar{\sigma}^{(1)}_{\bar{e} e}(s')}{s}  +\mathcal{O}\left(\frac{m_e^2}{s}\right),  
\end{eqnarray}
where 
\begin{eqnarray}
&& P_{ee}^{(0)} { (z)}= \left[\frac{1 +z^2}{1-z}\right]_+ , \\
&& d_{ee}^{(1)}(z)  = \left[\frac{1 +z^2}{1-z}\biggl(\ln \frac{\mu_R^2}{m_e^2}- 1 - 2 \ln(1-z) \biggr)\right]_+   
\end{eqnarray}
and "$+$" means plus-presctription, and $\bar{\sigma}^{(1)}_{\bar{e} e}$ is the one-loop contribution to the annihilation sub-process cross section for massless partons computed within the $\overline{\mathrm{MS}}$ scheme for subtraction of electron mass singularities. 
It is convenient to choose $\mu_R=m_e$ so that logarithms of different energy scale ratios will be collected in one place, i.e., to take the large logarithm being $L=\ln(\mu_F^2/m_e^2)$. It should also be noted that choosing $\mu_R = \mu_F$ will completely eliminate the benefit of using the structure function method in QED, since all large logs will become zero, and large contributions will shift to terms of order $\mathcal{O}(\alpha^nL^0)$, which are very difficult to calculate. Note also that the left hand side of Eq.~\eqref{matchingOa1} doesn't depend on $\mu_F$, so the dependence on this parameter cancels out on the right hand side between the first term in he square brackets and $\bar{\sigma}^{(1)}_{\bar{e} e}$. Moreover, cancellation of
the $\mu_F$-dependence happens in each line of Eq.~\eqref{triangle} (starting from the second one).

One can take $\mu_F^2 = s$ as the first option of the factorization scale choice so that $L=\ln(s/m_e^2)$. 
Another option is adopted in Refs.~\cite{Berends:1987ab} and \cite{Ablinger:2020qvo,Blumlein:2011mi} where $\mu_F^2 = s z$ is chosen, so the factorization scale is equal to the invariant mass of the final $\mu^+\mu^-$ system and the large logarithm is $\hat L = L  + \ln z$. Note that in the latter case the factorization scale is equal to the energy transfer of the hard sub-process, i.e., the CSS prescription is applied. 

The master equation \eqref{master} for the cross section can be expanded as
\begin{eqnarray} \label{cij}
	\frac{d\sigma^{\mathrm{NNLL}}_{\bar{e} e}(s')}{ds'}  =  \frac{\sigma^{(0)}_{\bar{e} e}(s')}{s} \biggl\{ 1 +  \sum \limits_{l=0,1} 
	c_{1l}(z) \left(\frac{\alpha}{2\pi}\right)  L^{l}  + \sum \limits^{\infty}_{\substack{k=2 \\ k\geq l \geq k-2}} 
	c_{kl}(z) \left(\frac{\alpha}{2\pi}\right)^k  L^{l}       
     + \mathcal{O}(\alpha^3 L^{0}) \biggr\},
\end{eqnarray}
Higher-order coefficients $c_{kl}(z)$ in the leading and next-to-leading logarithmic approximations up to $c_{65}$ were first calculated in~\cite{Ablinger:2020qvo} and some of those coefficients were independently recalculated and corrected in~\cite{Arbuzov:2024tac}. Explicit expression for additions to $c_{20}$ with respect to the result in Berends' work \cite{Berends:1987ab} can be found in~\cite{Blumlein:2019srk}. For convenience, present the explicit result in Appendix.

A change of the factorization scale and the subsequent change of the large logarithm leads to redistribution of contributions in each order in $\alpha$. 
Let us now consider radiative corrections of the $\mathcal{O}(\alpha^2)$ order. Any change of the factorization scale must not change the full correction in the given order in $\alpha$ (and the complete result for the cross-section):
\begin{equation}
h_{22} + h_{21} + h_{20} = \hat{h}_{22} + \hat{h}_{21} + \hat{h}_{20}, 
\end{equation}
where we denote 
\begin{equation}
h_{ij} \equiv \left( \frac{\alpha}{2 \pi} \right)^i L^j c_{ij}, \quad
  \hat{h}_{ij} \equiv \left( \frac{\alpha}{2 \pi} \right)^i \hat{L}^j \hat{c}_{ij}, \quad \hat{L} = \ln \frac{\hat{\mu}_F^2}{m_e^2}.
\end{equation}
We consider changing the factorization scale as an addition to the large logarithm:
\begin{equation}
    \hat{L} = L + \Delta L, \quad \Delta L = \ln \frac{\hat{\mu}_F^2}{{\mu}_F^2},
\end{equation}
 and we get the matching equality:
\begin{eqnarray}
\left( \frac{\alpha}{2 \pi} \right)^2 L^2 c_{22}  + \left( \frac{\alpha}{2 \pi} \right)^2 L c_{21} + \left( \frac{\alpha}{2 \pi} \right)^2 c_{20}  = \left( \frac{\alpha}{2 \pi} \right)^2 \hat{L}^2 \hat{c}_{22}  + \left( \frac{\alpha}{2 \pi} \right)^2 \hat{L} \hat{c}_{21} + \left( \frac{\alpha}{2 \pi} \right)^2 \hat{c}_{20} .
\end{eqnarray}

We get expressions for the new coefficients $\hat{c}_{ij}$:
\begin{eqnarray} \label{abd}
  &&  \hat{c}_{22} = c_{22}, \nonumber \\
  &&  \hat{c}_{21} = c_{21} - 2 \Delta L c_{22}, \nonumber \\
  &&  \hat{c}_{20} = c_{20} -  \Delta L  c_{21} + (\Delta L)^2 c_{22},
\end{eqnarray}

\begin{figure*}
\includegraphics[scale=2]{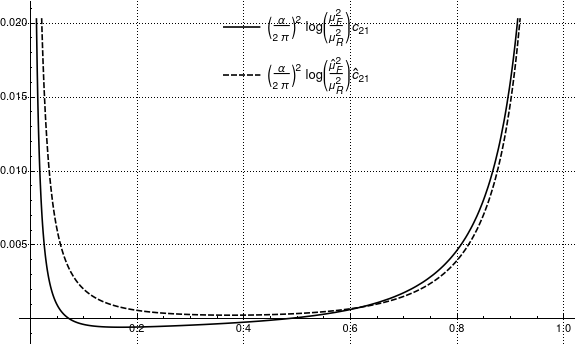}
\caption{\label{c21z} $\mathcal{O}(\alpha^2L^1)$ contributions for different factorization scales vs. energy fraction $z$}
\end{figure*}

\begin{figure*}
\begin{center}
\includegraphics[scale=2]{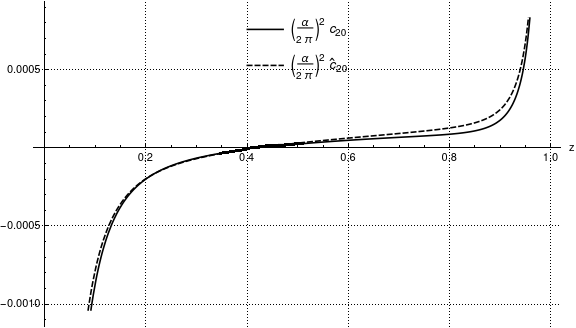}
\caption{\label{c20z} $\mathcal{O}(\alpha^2L^0)$ contributions for different factorization scales}
\end{center}
\end{figure*}

The behavior of functions $c_{ij}(z)$ for $\mu_F=\sqrt{s}$ and $\hat{c}_{ij}(z)$ for $\hat{\mu}_F=\sqrt{sz}$ in the $\mathcal{O}(\alpha^2)$ order is shown in Figs.~\ref{c21z} and \ref{c20z}. 
At $z\to 0$, the functions grow rapidly due to the presence of terms proportional to $1/z$ and $\ln(z)$, which are related to the singlet pair production mechanism. If necessary, the resummation of such terms can be performed within the Balitsky-Fadin-Kuraev-Lipatov (BFKL) formalism~\cite{Fadin:1975cb}.
The peaked behavior at $z\to 1$ is governed by soft photon and pair emission, and it is regularized by the corresponding terms due to virtual radiative corrections, for more details, see~\cite{Arbuzov:2024tac}.


\begin{figure*}
    \centering
    \includegraphics[width=0.95\linewidth]{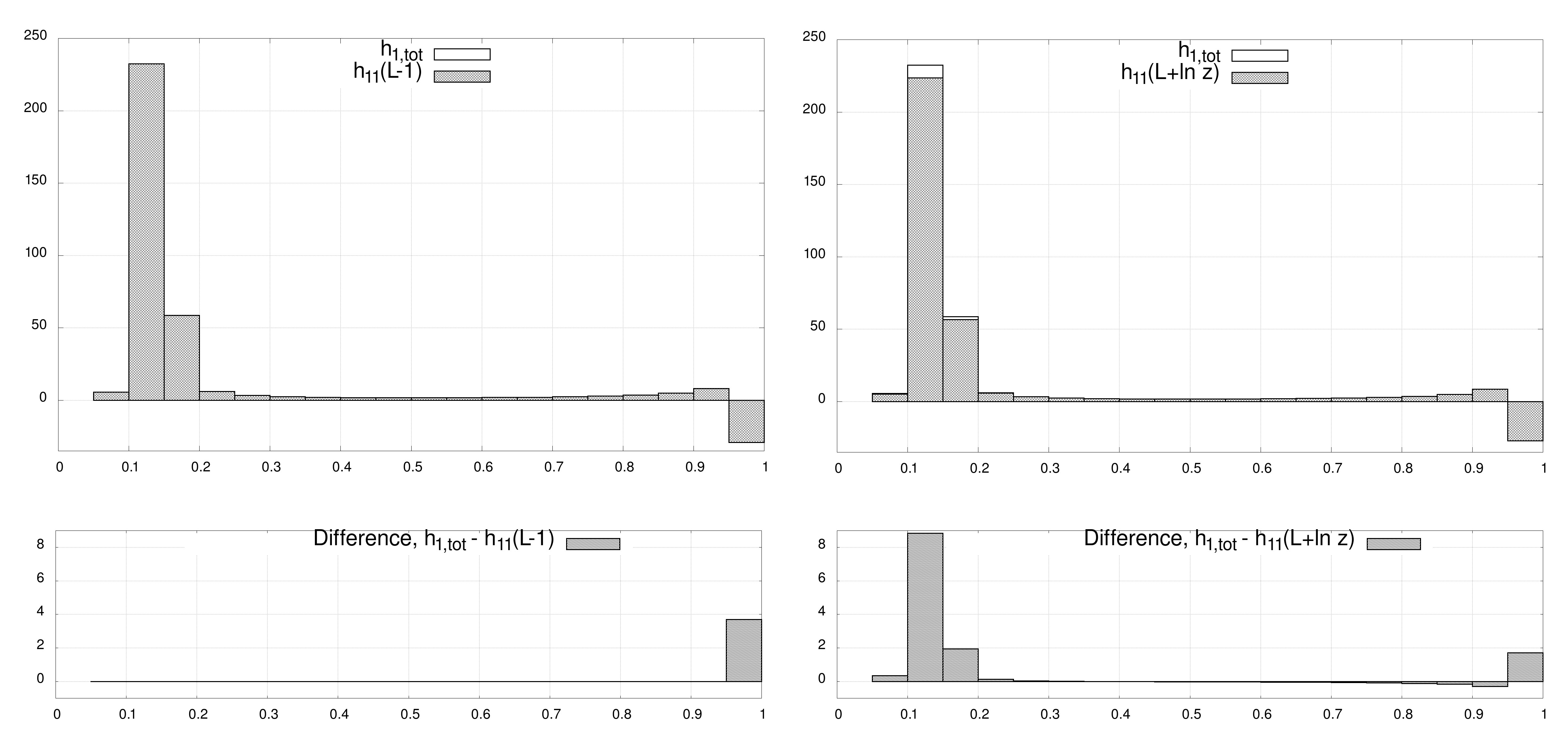}
    \caption{First-order LL corrections and their difference from the total
    LL+NLL result for two factorization scales $\sqrt{s/e}$ (left) and $\sqrt{s z}$ (right) at
$\sqrt{s}=240$~GeV, in $\%$, vs. energy fraction $z$}
    \label{Oa1}
\end{figure*}

\begin{figure*}
    \centering
    \includegraphics[width=0.95\linewidth]{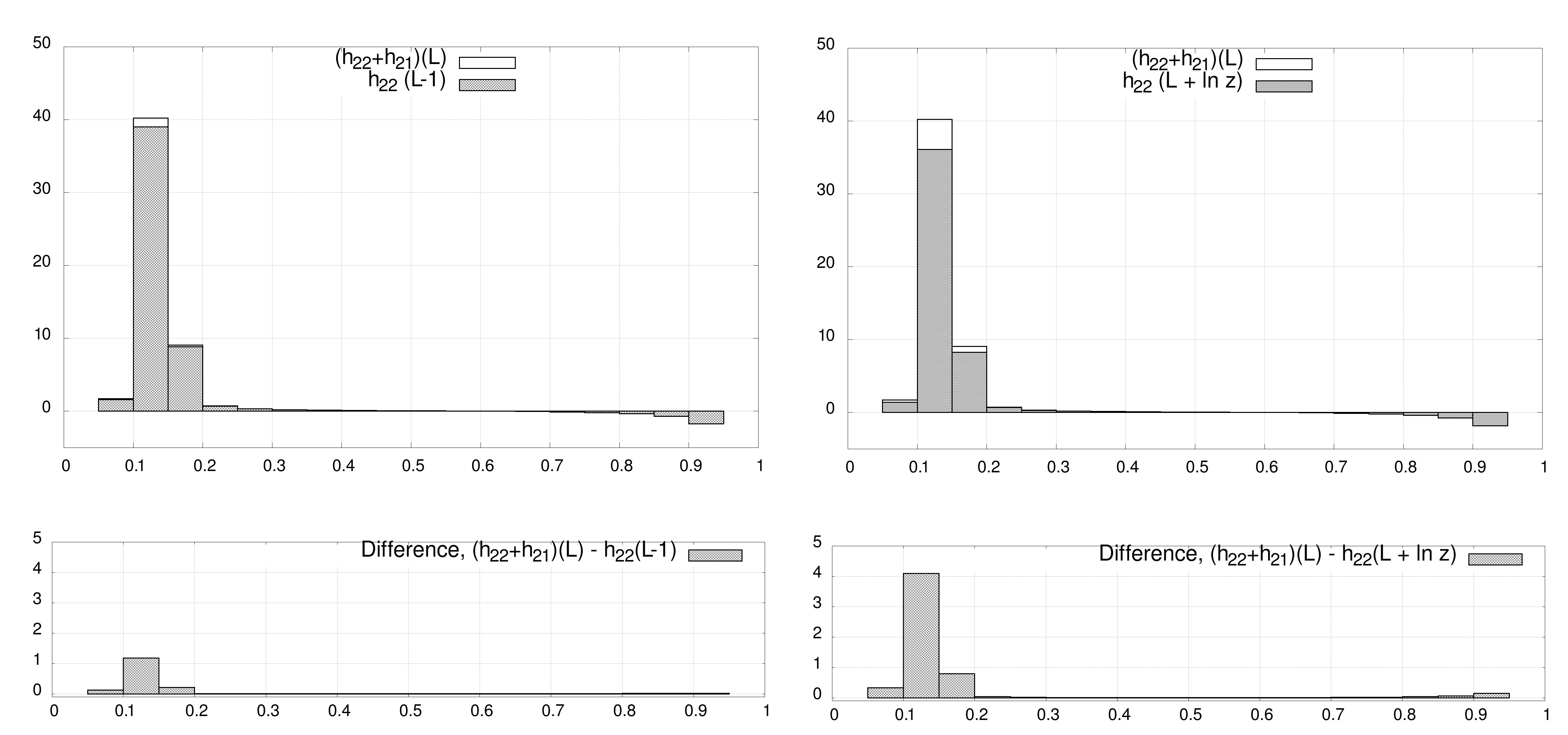}
    \caption{Second-order LL corrections at two factorization scales: $\sqrt{s/e}$ (left) and $\sqrt{s z}$ (right), and their difference from the
    LL+NLL result at $\mu_F=\sqrt{s}$ at $\sqrt{s}=240$~GeV, in $\%$, vs. energy fraction $z$}
    \label{Oa2NLL}
\end{figure*}

\begin{figure*}
    \centering
    \includegraphics[width=0.95\linewidth]{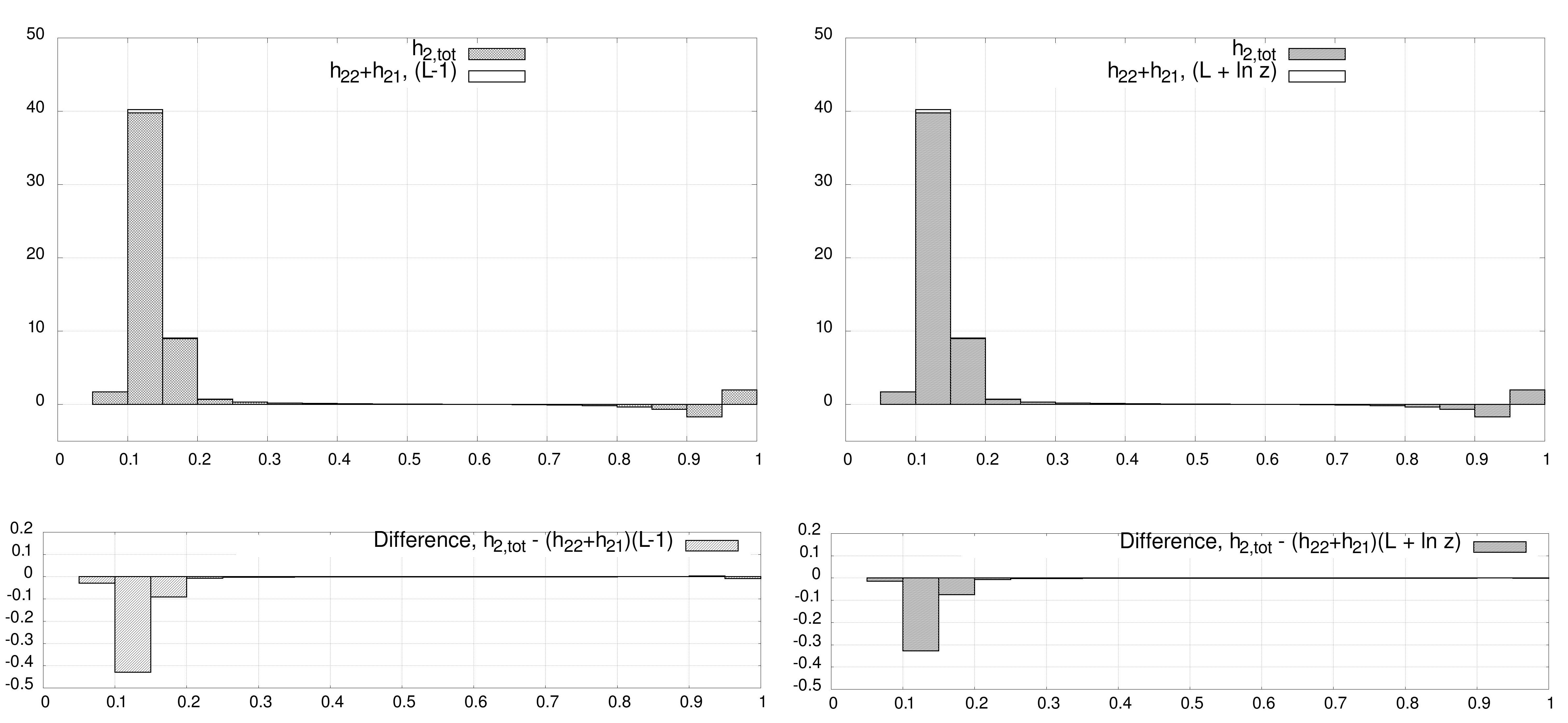}
    \caption{Second-order NLL corrections and their difference from the total
    NNLL result for two factorization scales $\sqrt{s/e}$ (left) and $\sqrt{s z}$ (right) at
$\sqrt{s}=240$~GeV, in $\%$, vs. energy fraction $z$}
    \label{Oa2NNLL}
\end{figure*}

At $\sqrt{s}=240$ GeV we have the radiative return to the $Z$ resonance at $z \approx 0.14$, where $sz=M_Z^2$. At $\sqrt{s}=160$~GeV the $Z$ resonance peak is at $z \approx 0.32$. 
Let us look at the contributions to the differential cross section of the orders $\mathcal{O}(\alpha)$ and $\mathcal{O}(\alpha^2)$ at $\sqrt{s} = 240$~GeV. 
The interval from $z=0$ to $z=1$ is divided into 20 bins and the integration was performed separately for each bin, starting from $z_{min} = 0.05$. The histogram for the cross section with two different factorization scales equal to $\sqrt{s/e}$ and $\sqrt{s z}$ (marked on the plot as "$(L-1)$" and "$(L+ \ln z)$") and the difference from the total result for the correction in the given order are shown in Figs.~\ref{Oa1} and \ref{Oa2NNLL}. In Fig.~\ref{Oa2NLL} the LL corrections at $\sqrt{s/e}$ and $\sqrt{s z}$  factorization scales are compared with LL+NLL contribution at the $\sqrt{s}$ factorization scale. This values are shown in percent relatively to the Born cross section $\sigma^{(0)}_{\bar{e}e}(s)$ \cite{Berends:1987ab}. Here $h_{i,tot}$ is a sum of all corrections of the given $i^{\mathrm{th}}$ order in $\alpha$, e.g.
\begin{equation}
h_{1,tot} = h_{11} + h_{10}.
\end{equation}
The preferred factorization scale appears to be $\sqrt{s/e}$, because with this scale in the order $\mathcal{O} (\alpha^1)$ the NLL contribution is almost completely (except the last bin) absorbed into LL one, see Fig.~\ref{Oa1}. A similar situation happens with the choice of the preferable factorization scale in the leading $\mathcal{O}(\alpha^2L^2)$ contribution which absorbs the bulk of 
$\mathcal{O}(\alpha^2L^1)$ effect if $\mu_F=\sqrt{s/e}$,
as one can see from Fig.~\ref{Oa2NLL}.
But when we go to the NLL level and look at the sum of the $\mathcal{O}(\alpha^2L^2)$ and $\mathcal{O}(\alpha^2L^1)$ corrections, we see in Fig.~\ref{Oa2NNLL} that for $\mu_F=\sqrt{sz}$ most NNLL contributions are absorbed into the NLL ones. Note that this happened for the integrated (over bins) distribution and for the particular choice of the c.m.s. energy. 

\subsection{Factorization scale choices in the total cross section}
By integrating the corrected cross section from $z_{min}=0.1$ to $1$ at $\sqrt{s}=240$~GeV, we include the contribution due to the radiative return to the $Z$ resonance into the integration domain, and that yields unusually large corrections.  Thus, at $e^+e^-$ colliders with energies above the $Z$ peak, one can expect quite a lot of events due to radiative return to the peak, and the theoretical uncertainties in the description of such events will suffer from large, yet unknown, higher-order contributions.

\begin{figure*}
    \centering
    \includegraphics[width=0.95\linewidth]{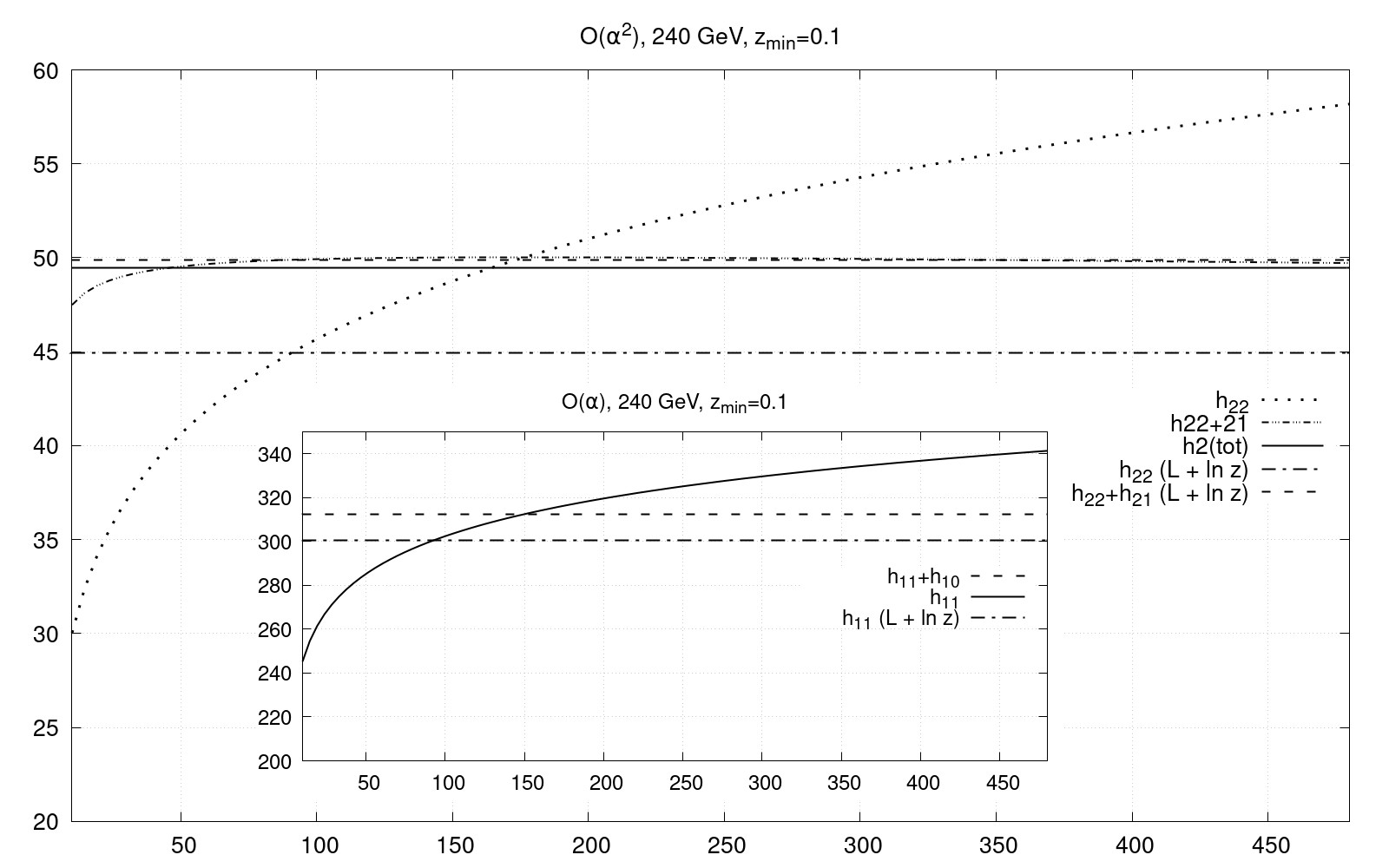}
    \caption{Corrections vs. factorization scale at $\sqrt{s}=240$ GeV, $z_{min}=0.1$}
    \label{fig:pms240}
\end{figure*}

\begin{figure*}
    \centering
    \includegraphics[width=0.95\linewidth]{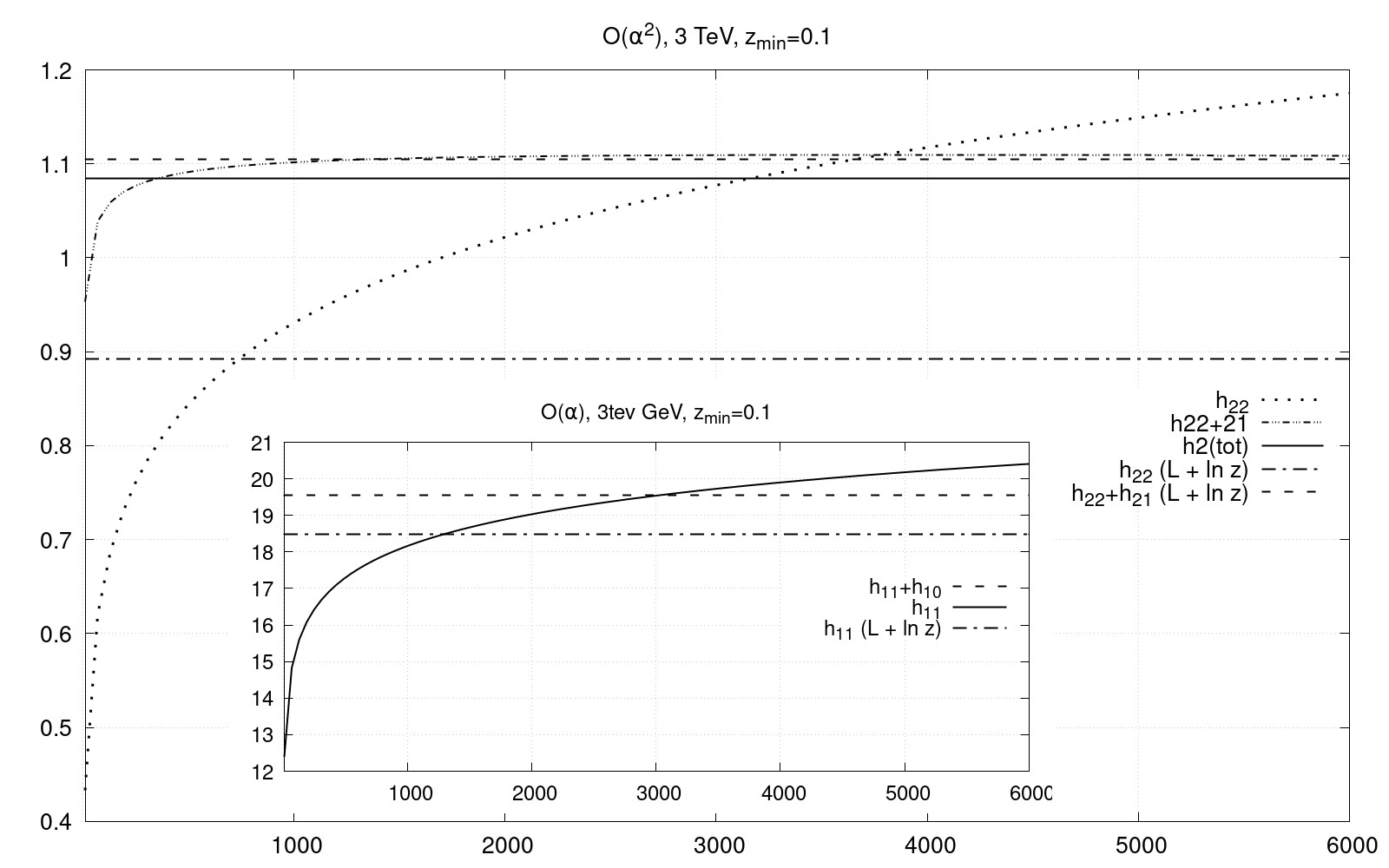}
    \caption{Corrections vs. factorization scale at 
    $\sqrt{s}=3$ TeV, $z_{min}=0.1$}
    \label{fig:pms5K}
\end{figure*}

Figs. \ref{fig:pms240} and \ref{fig:pms5K} show the LL and NLL corrections as a function of the factorization scale choice with $z_{min}=0.1$. They are compared to the results with the factorization scale $\mu_F = \sqrt{sz}$, the corresponding lines are marked $h_{ij} (L + \ln z)$. The center-of-mass (c.m.s.) energies are $240$~GeV and $3$~TeV, which correspond to the optimal energy of associated production of Higgs and $Z$ bosons at FCC-ee and CEPC colliders and to the maximal energy of the CLIC collider project~\cite{Linssen:2012hp}, respectively. The factorization scale varies from $10$~GeV to the maximal c.m.s. energy. The complete correction of the particular order in $\alpha$ does not depend on the factorization scale. Figs.~\ref{fig:pms240} and \ref{fig:pms5K} show the $\mathcal{O}(\alpha^1)$ and $\mathcal{O}(\alpha^2)$ corrections with different scaling for better readability.
One can see that the leading logarithmic contributions $h_{11}$ and $h_{22}$ strongly depend on
$\mu_F$, while in the next-to-leading approximation the dependence becomes rather weak. The conventional 
choice of the factorization scale $\mu_F = \sqrt{sz}$ is certainly not optimal in the LL approximation,
but occasionally it works reasonably well in the NLL case.  According to principle of minimal sensitivity, the best factorization scale is where the derivative of the function is equal to zero. This approach doesn't work for the leading logarithmic approximation since there the dependence on $\mu_F$ is monotonic.
For $h_{22}+h_{21}$ there is a maximum, which is situated at $\mu_F \approx 175.6$ GeV for $\sqrt{s}=240$ GeV and at $\mu_F \approx 3.8$ TeV for $\sqrt{s}=3$ TeV. But the second-order correction $h_{22}+h_{21}$ is almost flat, so the PMS approach doesn't appear to be much instructive (especially if there would be additional uncertainties, e.g., due to non-perturbative effects in QCD). The FAC approach provides a clear simple recipe to choose the factorization scale at the intersection of the curves corresponding to the leading and next-to-leading approximations. This scale choice actually provides a quite reasonable approximation of the total $O(\alpha^2)$ result.

Note that the difference between the NLL contribution $h_{22}+h_{21}$ and the total $O(\alpha^2)$ result is almost insensitive to the factorization scale choice. It means that this difference hardly can be absorbed into the factorization scale choice. The reason is that this difference is mainly coming from the contributions which are not proportional the ones enhanced by the large logarithms. In particular, we see that 
BFKL-like terms proportional to $\ln^k(z)/z, \ \ k\geq 0$ are important numerically when small $z$ values are not forbidden by experimental cuts.

It looks reasonable to choose the factorization scale for at the intersection of the LL and LL+NLL correction curves. This 
choice corresponds to the fastest apparent convergency prescription.  At $\sqrt{s}=240$~GeV for the order $\mathcal{O(\alpha)}$, the corresponding scale is about $150$~GeV, and one can see that it is close to $\sqrt{s/e}=  {240}/\sqrt{e}\approx146$~GeV.
It is important that in higher orders, due to more complicated structure of corrections, this intersection point is shifted. Moreover, the position of the intersection point depends on experimental cuts, e.g. on $z_{min}$.  Thus, there can be no ideal choice of factorization scale but its optimization is nonetheless very important.

The numerical values of the corrections with different $z_{min}$ are given in the Table~\ref{Tab.Rel}. We present there the full $\mathcal{O}(\alpha^2)$ corrections, the sum of the $\mathcal{O}(\alpha^2L^2)$ and $\mathcal{O}(\alpha^2L^1)$ contributions and separately the $\mathcal{O}(\alpha^2L^2)$ part for different factorization scale values. Corrections are significantly bigger at $z_{min}=0.1$ than at $z_{min}=0.5$ due to the radiative return to the resonance being allowed in the first case. The corrections also increase at $z_{min}=0.9$ due to the soft radiation contributions. The full $\mathcal{O} (\alpha^2)$ correction, as it should be, doesn't depend on the factorization scale value. One can see that for $\mu_F=\sqrt{s/e}$, the LL result $h_{22}$ is more close to the total sum $h_{22}+h_{21}+h_{20}$ except the case $z_{min}=0.1$ at $\sqrt{s}=3$~TeV where $\mu_F=\sqrt{s}$ works better. Meanwhile, in the NLL approximation $\mu_F=\sqrt{sz}$ works somewhat better. It can also be noted that for $z_{min}=0.1$, the variation in the factorization scale among the three values under consideration seriously underestimates the difference between the NLL and NNLL results.

\begin{table*}[ht] 
\caption{Corrections of the order $\mathcal{O}(\alpha^2)$ in \% to the total cross section for $\sqrt{s} = 240$~GeV and $3$ TeV at different $z_{min}$ } 
       \label{Tab.Rel}
         \begin{tabular}{|l|c|c|c|c|c|c|c|} 
          \hline 
          \multicolumn{2}{|c|}{} & \multicolumn{3}{|c|}{$240$ GeV} &  \multicolumn{3}{|c|}{$3$ TeV} \\     
            \hline
            \multicolumn{2}{|c|}{} & $z_{min} = 0.1$ & $z_{min} = 0.5$ & $z_{min} = 0.9$ & $z_{min} = 0.1$ & $z_{min} = 0.5$ & $z_{min} = 0.9$ \\
		\hline
		  \multicolumn{2}{|c|}{$h_{22} + h_{21} + h_{20}$} &  $49.4704$ & $-0.8604$ & $0.3044$ & $1.0844$ & $-1.1743$ & $0.5282$\\
             \hline
              \multirow{3}{4em}{$h_{22} + h_{21}$} & $\sqrt{s}$ & $ 49.9954$ & $- 0.8812 $ & $0.3063$ & $1.1092$ & $-1.1656$ & $0.5302$ \\
              &$\sqrt{sz}$ & $49.8924$ & $-0.8838 $ & $0.3052$ & $1.1044$ & $-1.1683$ & $0.5291$ \\
              & $\sqrt{s/e}$ & $ 50.0130 $ & $-0.8841$ & $0.3086$ & $1.1071$ & $-1.1682$ & $0.5325$ \\
                           \hline         
              \multirow{3}{4em}{$h_{22}$} & $\sqrt{s}$  & $52.4635$ & $- 0.9954$ & $0.3793$ & $1.0772$ & $-1.2869$ & $0.6229$ \\
              & $\sqrt{sz}$  & $44.9544$ & $-0.9508$ & $0.4077$ & $0.8922$ & $-1.2342$ & $0.6565$ \\
              & $\sqrt{s/e}$  & $48.5231$ & $-0.9206$ & $0.3508$ & $1.0092$ & $-1.2057$ & $0.5836$ \\
              \hline
		\end{tabular}
\end{table*}

\subsection{Uncertainties due to scale variation}

Let us now look at the effect due to variation of the factorization scale by factor 2. The magnitude of the corresponding effect 
is commonly used to estimate unaccounted higher-order corrections~\cite{Maltoni:2007tc}.

We choose the central value of the factorization scale $\mu_F$ and compute $h_{22}$, $h_{22} + h_{21}$, and $h_{33}$ at factorization scales $2 \mu_F$ and $\mu_F/2$. 
In Eq. \eqref{delta_def}  $\Delta_2^{LL}$ and $\Delta_3^{LL}$ are $h_{21}$ and $h_{32}$ respectively, and $\Delta_2^{NLL}$ is $h_{20}$ at the central value of the factorization scale. So, $\Delta_2^{LL}$ and $\Delta_3^{LL}$, and $\Delta_2^{NLL}$  are the true values of the shifts due to "unknown contributions" if one works in the LL and NLL approximations, respectively. Quantities $\delta^{LL}$ and $\delta^{NLL}$ show the average shifts of the LL and NLL contributions due to the factorization scale variation. 
So, $\delta_{LL,NLL}$ are estimated shifts of the corresponding "unknown contributions". The true and estimated shifts are defined as follows:
\begin{eqnarray} \label{delta_def}
	&& \Delta^{\mathrm{LL}}_2 = |h_{21}|, \nonumber \\
	&& \Delta^{\mathrm{NLL}}_2 = |h_{20}|, \nonumber \\
	&& \delta^{\mathrm{LL}}_2 = \frac{|h_{22} - h_{22} (1/2)| + |h_{22} - h_{22} (2)|}{2} , \nonumber \\
	&& \delta^{\mathrm{NLL}}_2 = \frac{|h_{22}+h_{21} - (h_{22}+h_{21}) (1/2)|}{2}  \nonumber \\ 
    && + \frac{|h_{22}+h_{21} - (h_{22}+h_{21}) (2)|}{2}, \nonumber \\
    && \Delta^{\mathrm{LL}}_3 = |h_{32}|, \nonumber \\
    && \delta^{\mathrm{LL}}_3 = \frac{|h_{33} - h_{33} (1/2)| + |h_{33} - h_{33} (2)|}{2},
\end{eqnarray}
where $h_{ij}(2)$ and $h_{ij}(1/2)$ are $h_{ij}$ at factorization scales $2 \mu_F$ and $\mu_F/2$ correspondingly, and $h_{ij}$ without argument are $h_{ij}$ at initial (central) factorization scale. Numerical values of $h_{ij}$ can be found in our work \cite{asymmAV:2026}.

The numerical results for the shifts are given in Tables~\ref{Tab.delta_s}---\ref{Tab.delta_sz} for three different values of the factorization scale and three energies.
It is seen that the average shift in the LL is of the same order as the corresponding NLL contribution ($h_{21}$ or $h_{32}$), but in the NLL at $\sqrt{s} = M_z$ the variation of the factorization scale significantly overestimates $h_{20}$ for $\mu_F=\sqrt{s}$ and $\mu_F=\sqrt{sz}$. At $\sqrt{s} = M_z$ both in $\mathcal{O}(\alpha^2)$ and $\mathcal{O}(\alpha^3)$ orders $\Delta$ and $\delta$ almost don't change with the change of $z_{min}$, since radiative return in this case is not efficient.
However, at the energies above the $Z$ peak (for $\sqrt{s}=240$~GeV or 3~TeV) for $z_{min}=0.1$
we see that the estimated shifts $\delta_2^{\mathrm{NLL}}$ and $\delta_3^{\mathrm{LL}}$ seriously underestimate the corresponding true values. Note that for $\sqrt{s}=240$~GeV at $z_{min}=0.1$ the radiative return to the resonance 
is taken into account, while for $\sqrt{s}=3$~TeV it contributes only for $z_{min}<0.001$. So the radiative return happens just to lower center-of-mass energies where the cross-section is higher. For both $\mu_F=\sqrt{sz}$ and $\mu_F=\sqrt{s/e}$ the variation underestimates the real NLL correction at $\sqrt{s} = 3$~TeV and $z_{min}=0.1$ because of QED BFKL effects at small $z$ values\footnote{A detailed analysis of these effects will be presented elsewhere.}. Both these scales are not optimal for $3$~TeV. Better scale choice here can be given by the FAC and PMS approaches ($\mu_F\approx 3.8$~TeV).

	\begin{table*}[ht] 
		\caption{Calculated and estimated uncertainties in $\%$ with central value $\mu_F=\sqrt{s}$} \label{Tab.delta_s}
		\begin{tabular}{|l|c|c|c|c|c|c|} 
                \hline
                &  \multicolumn{4}{|c|}{$\mathcal{O}(\alpha^2)$} &  \multicolumn{2}{|c|}{$\mathcal{O}(\alpha^3)$} \\
			\hline
		  &  \multicolumn{2}{|c|}{LL} &  \multicolumn{2}{|c|}{NLL} &  \multicolumn{2}{|c|}{LL} \\
						\hline 
			 & $ \Delta_2^{\mathrm{LL}} $ & $\delta_2^{\mathrm{LL}} $  & $ \Delta_2^{NLL} $ & $\delta_2^{\mathrm{NLL}} $& $ \Delta_3^{\mathrm{LL}} $ & $\delta_3^{\mathrm{LL}} $ \\  
			\hline 
			$\sqrt{s}=M_z$, 
				& $0.436 $	& $0.524  $&	$0.0063 $ &$	0.0250  $ & $0.0250$ & $0.0499$
			 \\
			 			 $z_{min}=0.1$ & & & & & &\\
			 \hline
			 $\sqrt{s}=M_z$, 
			 & $0.436$&	$0.5246$	& $0.0062$ &	$0.0250$ & $0.0249$& $0.0499$
		 
			  \\
			 			  $z_{min}=0.5$ & & & & & &\\
			 \hline
			$\sqrt{s}=M_z$,  & $0.440$ &	$0.529$ &	$0.0062$ &	$0.0252$ &$0.0249$ & $0.0497$
			
			\\
						$z_{min}=0.9$ & & & & & &\\
			\hline
			$\sqrt{s}=240$ GeV, 	& $2.468$ &	$5.569$&	$0.5250$&	$0.1479$ & $0.5833$& $0.1015$
			
			\\
					$z_{min}=0.1$ & & & & & &\\
			\hline 
			$\sqrt{s}=240$ GeV, 
				&$0.1142$&	$0.1057$ &	$0.0092$ &$	0.0061$& $0.0014$& $0.0002$
				
			\\
				$z_{min}=0.5$ & & & && & \\
			\hline 
			$\sqrt{s}=240$ GeV,	&  $0.073$ &	$0.0403 $&$	0.0019$	& $0.0039$& $0.0216$& $0.0294$
						 \\
			$z_{min}=0.9$ & & & & & &\\
			\hline
            			$\sqrt{s}=3$ TeV, & $0.0320$ &	$0.0958$ & $0.0240$ & $0.0248$ & $0.0307$ & $0.0126$
			\\
					$z_{min}=0.1$ & & & & & &\\
			\hline 
			$\sqrt{s}=3$ TeV, 
				& $0.1213$ &  $0.1145$ & $0.0082$ & $0.0087$  & $0.0034$ & $0.0018$
				
			\\
				$z_{min}=0.5$ & & & & & & \\
			\hline 
			$\sqrt{s}=3$ TeV, & $0.0926$ & $0.0554$ & $0.0020$ & $0.0041$ & $0.0298$ & $0.0402$
						 \\                 
			$z_{min}=0.9$ & & & & & &\\
			\hline
	\end{tabular}
	\end{table*}

	\begin{table*}[ht] 
		\caption{Calculated and estimated uncertainties in $\%$ with central value $\mu_F=\sqrt{s/e}$} \label{Tab.delta_se}
		\begin{tabular}{|l|c|c|c|c|c|c|} 
                \hline
                &  \multicolumn{4}{|c|}{$\mathcal{O}(\alpha^2)$} &  \multicolumn{2}{|c|}{$\mathcal{O}(\alpha^3)$} \\
			\hline
			  &  \multicolumn{2}{|c|}{LL} &  \multicolumn{2}{|c|}{NLL} &  \multicolumn{2}{|c|}{LL} \\
						\hline 
			 & $ \Delta_2^{\mathrm{LL}} $ & $\delta_2^{\mathrm{LL}} $  & $ \Delta_2^{NLL} $ & $\delta_2^{\mathrm{NLL}} $& $ \Delta_3^{\mathrm{LL}} $ & $\delta_3^{\mathrm{LL}} $ \\  
			\hline 
			$\sqrt{s}=M_z$, 
				& $0.0556 $	& $0.5032 $&	$ 0.0165$ & $0.0150$ & $0.0101$ & $0.0459$

			 \\
			 			 $z_{min}=0.1$ & & & & & &\\
			 \hline
			 $\sqrt{s}=M_z$, 
			 &$ 0.0557 $ & $0.5030$	& $ 0.0165$ &	$0.0150$ & $ 0.0101$& $ 0.0459$

			  \\
			  			  $z_{min}=0.5$ & & & & & &\\
			 \hline
			$\sqrt{s}=M_z$,  & $0.0567 $ &	$ 0.5067$ &	$0.0166 $ &	$ 0.0152$ &$0.0101 $ & $ 0.0457$
			
			\\
						$z_{min}=0.9$ & & & & & &\\
			\hline
			$\sqrt{s}=240$ GeV, 	& $1.4898 $ &	$ 5.3558$&	$ 0.5426 $&	$0.1478$ & $0.4718$& $ 0.0939$
			
			\\
					$z_{min}=0.1$ & & & & & &\\
			\hline 
			$\sqrt{s}=240$ GeV, 
				&$ 0.0366 $&	$0.1016$ &	$0.0063$ &$0.0028 $& $0.0011 $& $0.0002$
				
			\\
				$z_{min}=0.5$ & & & && & \\
			\hline 
			$\sqrt{s}=240$ GeV,	&  $0.0423$ &	$0.0387$ & $0.0042$	& $0.0023$ & $0.0004 $ & $0.0271$
						 \\
			$z_{min}=0.9$ & & & & & &\\
			\hline
            			$\sqrt{s}=3$ TeV, & $0.0979 $ &	$ 0.0927$ & $0.0227$ & $0.0045$ & $ 0.0202$ & $ 0.0118$
			\\
					$z_{min}=0.1$ & & & & & &\\
			\hline 
			$\sqrt{s}=3$ TeV, 
				& $0.0374$ &  $ 0.1108$ & $0.0061$ & $0.0025$  & $0.0020$ & $0.0017$
				
			\\
				$z_{min}=0.5$ & & & & & & \\
			\hline 
			$\sqrt{s}=3$ TeV, & $0.0510$ & $0.0536$ & $0.0043$ & $0.0023$ & $0.0007$ & $0.0377$
						 \\
			$z_{min}=0.9$ & & & & & &\\
			\hline
		\end{tabular}
	\end{table*}

	\begin{table*}[ht] 
		\caption{Calculated and estimated uncertainties in $\%$ with central value $\mu_F=\sqrt{sz}$} \label{Tab.delta_sz}
		\begin{tabular}{|l|c|c|c|c|c|c|} 
                \hline
                &  \multicolumn{4}{|c|}{$\mathcal{O}(\alpha^2)$} &  \multicolumn{2}{|c|}{$\mathcal{O}(\alpha^3)$} \\
			\hline
			  &  \multicolumn{2}{|c|}{LL} &  \multicolumn{2}{|c|}{NLL} &  \multicolumn{2}{|c|}{LL} \\
						\hline 
			 & $ \Delta_2^{\mathrm{LL}} $ & $\delta_2^{\mathrm{LL}} $  & $ \Delta_2^{NLL} $ & $\delta_2^{\mathrm{NLL}} $& $ \Delta_3^{\mathrm{LL}} $ & $\delta_3^{\mathrm{LL}} $ \\  
			\hline 
			$\sqrt{s}=M_z$, 
				& $0.4491$	& $0.5256$&	$0.0058$ & $0.0257$ & $ 0.0281$ & $0.0503$

			 \\
			 			 $z_{min}=0.1$ & & & & & &\\
			 \hline
			 $\sqrt{s}=M_z$, 
			 & $0.4492$ & $0.5254$	& $0.0058$ &	$0.0257$ & $ 0.0281$& $ 0.0503$

			  \\
			  			  $z_{min}=0.5$ & & & & & &\\
			 \hline
			$\sqrt{s}=M_z$,  & $0.4527 $ &	$0.5292$ &	$0.0057$ &	$0.0259$ &$0.0280$ & $0.0500$
			
			\\
						$z_{min}=0.9$ & & & & & &\\
			\hline
			$\sqrt{s}=240$ GeV, 	& $4.9380$ &	$ 5.1551$&	$0.4220 $&	$0.2829$ & $0.3848$& $0.0877 $
			
			\\
					$z_{min}=0.1$ & & & & & &\\
			\hline 
			$\sqrt{s}=240$ GeV, 
				&$0.0670$&	$ 0.1033$ &	$0.0065$ &$0.0037$& $0.0039 $& $0.0005$
				
			\\
				$z_{min}=0.5$ & & & && & \\
			\hline 
			$\sqrt{s}=240$ GeV,	&  $0.1024$ &	$0.0418$ & $0.0008$	& $0.0054 $ & $0.0161 $& $0.0288$
						 \\
			$z_{min}=0.9$ & & & & & &\\
			\hline
            			$\sqrt{s}=3$ TeV, & $ 0.2121$ &	$0.0873$ & $0.0201$ & $0.0099$ & $0.0142$ & $0.0114$
			\\
					$z_{min}=0.1$ & & & & & &\\
			\hline 
			$\sqrt{s}=3$ TeV, 
				& $0.0659$ &  $0.1121$ & $0.0061$ & $0,0030$  & $0.0060$ & $0.0020$
				
			\\
				$z_{min}=0.5$ & & & & & & \\
			\hline 
			$\sqrt{s}=3$ TeV, & $0.1274$ & $0.0569$ & $0.0009$ & $0.0056$ & $0.0221$ & $0.0396 $
						 \\
			$z_{min}=0.9$ & & & & & &\\
			\hline
		\end{tabular}
	\end{table*}
    
\section{Discussion}

We considered various options for choosing the factorization scale for the correction to the $e^+e^- \rightarrow \gamma^*/Z^*$ processes. Knowing high-precision theoretical predictions for them is needed for studies at future electron-positron colliders including FCC-ee and CEPC. We observe that the factorization scale dependence provides a considerable source of uncertainty in estimates of higher-order corrections due to initial state radiation. 
Having examined known complete results in the $\mathcal{O}(\alpha)$ and $\mathcal{O}(\alpha^2)$ orders, we conclude that in this process, high values of the factorization scale of the order of the initial center-of-mass energy $\mu_F\approx \sqrt{s}$ are preferable. This choice is in accord with the prescription provided by the fastest apparent convergence method. Application of the principle of minimal sensitivity also favors large $\mu_F \sim \sqrt{s}$. The conventional scale setting scheme yields $\mu_F=\sqrt{zs}$ which is the momentum (energy) transfer in the hard subprocess. This scale choice completely fails in the leading logarithmic approximation, but works sufficiently well in NLL for the considered beam energies. It should also be noted that the BLM and PMC prescriptions are not applicable in our case since we have the bulk of scale dependence coming not from the running coupling constant but from the cusp anomalous dimensions.

We also verified the validity of the standard scale uncertainty estimation procedure by varying it by a factor of 2. We found that the procedure provides reasonable numerical estimates for some setups, but not for the crucial case of radiative return to a resonance or just to lower energies of the hard subprocess. In the latter case, the standard method considerably underestimates the uncertainty.

Our results are relevant for constructing an advanced treatment of higher-order QED corrections for electron-positron annihilation processes studied at current and future colliders. 
Our study was conducted for the case of pure QED,
but the latter can be viewed as an Abelian reduction of the Drell-Yan process in QCD, where the issue of scale dependence is also quite important.
Therefore, some of the presented insights on scale dependence may also be relevant for QCD, in particular, for description of Drell-Yan-like processes at LHC. 

\section{Appendix}
The explicit expression for NNLL correction in the $\mathcal{O}(\alpha^2L^0)$ order for $\mu_F=\sqrt{s}$ collected from the results given in refs.~\cite{Berends:1987ab,Blumlein:2019srk,Hamberg:1990np} reads
\begin{eqnarray}
&& c_{20}(z) =  \frac{701}{18} - \frac{32}{9 (1-z)^3} + \frac{64}{(9 (1-z)^2} + \frac{352}{(9 (1-z)} - \frac{40}{27 z} - \frac{1229 z}{18} - \frac{163 z^2}{54} + \frac{32}{3 (1 + z)^2} \nonumber \\
&& \quad - \frac{16}{1 + z} + \zeta_3 \left(32 - \frac{40}{1-z} - \frac{56}{z} + 6 z\right) + 
 \zeta_2 \left(\frac{38}{3} - \frac{16}{3 (1-z)} - \frac{8}{3 z} + 34 z + \frac{8 z^2}{3} \right)  \nonumber \\
&& \quad + \Biggl(-\frac{506}{9} + \frac{64}{9 (1-z)} + 40 z + \frac{64 z^2}{9} \Biggr) \ln (1-z) + \Biggl( \frac{1}{3} + \frac{16}{3 (1-z)} + \frac{16}{3 z} - \frac{5 z}{3} - \frac{16 z^2}{3} \Biggr) \nonumber \\
&& \quad \times \ln^2 (1-z)  + \left(- \frac{10}{3 (1-z)} - \frac{4 z}{3} \right) \ln^3 z + \zeta_2 \ln(1+z)\left(-24 - \frac{24}{z} - 12 z \right)  \nonumber \\
&& \quad + \left(24 + \frac{24}{z} + 12 z\right) \ln z \ln^2(1+z) + \ln^2 z \Biggl[3 - \frac{8}{3 (1-z)^3} + \frac{32}{3 (1-z)^2} - \frac{50}{3 (1-z)} - \frac{4}{z}  \nonumber \\
&& \quad - \frac{212 z}{33} + \frac{4 z^2}{3} + \Bigl(-22 + \frac{28}{1-z} - \frac{8}{z} - 18 z \Bigr) \ln (1-z) + \left(20 + \frac{20}{z} + 10 z\right) \ln (1 + z) \Biggr] \nonumber \\
&& \quad + \ln z \Biggl[ \frac{605}{9} + \frac{80}{9 (1-z)^3} + \frac{80}{9 (1-z)^2} - \frac{322}{9 (1-z)}  - \frac{32}{9 (1 - z)^4} - \frac{40}{9 z} - \frac{328 z}{9} - \frac{64 z^2}{9} \nonumber \\
&& \quad + \frac{32}{3 (1 + z)^3} - \frac{64}{3 (1 + z)^2} - \frac{40}{1 + z} + \zeta_2 \left(44 - \frac{48}{1-z} + \frac{16}{z} + 18 z\right)  + \Biggl(- \frac{16}{3} - \frac{88}{3 (1-z)}  \nonumber \\
&& \quad - \frac{16}{3 z} - \frac{28 z}{3} + \frac{32 z^2}{3} \Biggr) \ln (1-z) + (8 + 8 z) \ln^2 (1-z) + \Bigl(20 + \frac{32}{1-z} + \frac{16}{3 z}  + 4 z - \frac{16 z^2}{3}\Bigr) \nonumber \\
&& \quad \times \ln(1 + z) + \left(-72 - \frac{72}{z} - 36 z \right) \ln^2 (1 + z) \Biggr] + \Biggl[-6 - \frac{70}{3 (1-z)} - \frac{8}{3 z} - \frac{82 z}{3} + 8 z^2  \nonumber \\
&& \quad  + \Bigl(20 - \frac{16}{1-z} + 20 z \Bigr) \ln (1-z) + \left(-10 + \frac{40}{1-z} - \frac{16}{z} - 24 z \right) \ln z \Biggr] \mathrm{Li}_2(1-z) \nonumber \\
&& \quad  + \Biggl[20 + \frac{32}{1-z}  + \frac{16}{3 z} + 4 z - \frac{16 z^2}{3} + \left(-8 - \frac{16}{1-z} + \frac{40}{z} + 8 z \right) \ln z \nonumber \\
&& \quad + \left(-48 - \frac{48}{z} - 24 z\right) \ln (1 + z) \Biggr] \mathrm{Li}_2( -z)  + \left(-48 - \frac{48}{z} - 24 z\right) \ln(1+z) \mathrm{Li}_2(1 + z) \nonumber \\
&& \quad  + \left(-38 + \frac{24}{1-z} + \frac{8}{z} - 20 z\right) \mathrm{Li}_3(1-z)  + \biggl(56 + \frac{32}{1-z} - \frac{40}{z} + 4 z \biggr) \mathrm{Li}_3(-z)\nonumber \\
&& \quad  + \biggl( -32 + \frac{64}{1-z} - \frac{16}{z} - 24 z \biggr) \mathrm{Li}_3(z) + \left(48 + \frac{48}{z} + 24 z \right) \mathrm{Li}_3(1 + z).
\end{eqnarray}
Note that this result differs from the one provided in \cite{Blumlein:2020jrf}.

%
%

\subsection*{Acknowledgment}
We are grateful to A.L.~Kataev for fruitful discussions.




\bibliographystyle{unsrt}
\bibliography{radcorr}

\end{document}